\documentclass[aps,prd,nofootinbib,showpacs,floatfix,superscriptaddress,preprintnumbers]{revtex4}
\usepackage{amsmath}
\usepackage{amssymb}
\usepackage{epsfig}
\usepackage{graphicx}
\usepackage{pstricks}
\usepackage{pst-coil}
\DeclareMathAlphabet{\mathsc}{OT1}{cmr}{m}{sc}
\newcommand {\ignore}[1]{}

\newcommand{\nn}  {\nonumber}

\def\10{$SO(10)$}
\def\21{SU(2) $\otimes$ U(1) }

\def\422{$SU(4) \otimes SU(2) \otimes SU(2)$}
\def\321{SU(3) $\otimes$ SU(2) $\otimes$ U(1)}
\def\gsim{\raise0.3ex\hbox{$\;>$\kern-0.75em\raise-1.1ex\hbox{$\sim\;$}}}
\def\lsim{\raise0.3ex\hbox{$\;<$\kern-0.75em\raise-1.1ex\hbox{$\sim\;$}}}

\def\lsim{\raise0.3ex\hbox{$\;<$\kern-0.75em\raise-1.1ex\hbox{$\sim\;$}}}
\def\gsim{\raise0.3ex\hbox{$\;>$\kern-0.75em\raise-1.1ex\hbox{$\sim\;$}}}
\def\vev#1{\left\langle #1\right\rangle}
\def \znbb {0\nu\beta\beta}

\newcommand{\AddrAHEP}{%
  AHEP Group, Institut de F\'{\i}sica Corpuscular --
  C.S.I.C./Universitat de Val{\`e}ncia \\
  Edificio Institutos de Paterna, Apt 22085, E--46071 Valencia, Spain}

 \newcommand{\ba}{\begin{array}}
\newcommand{\ea}{\end{array}}
\relax

\def\321{$SU(3)\times SU(2)\times U(1)$}

\begin{document}

\preprint{IFIC/11-19}

\renewcommand{\Huge}{\Large}
\renewcommand{\LARGE}{\Large}
\renewcommand{\Large}{\large}
\def \znbb {$0\nu\beta\beta$ }
\def \nbb {$\beta\beta_{0\nu}$ }
\title{Admixture of quasi-Dirac and Majorana neutrinos with tri-bimaximal mixing}  \date{\today}
\author{S. Morisi} \email{morisi@ific.uv.es} \affiliation{\AddrAHEP}
\author{E. Peinado} \email{epeinado@ific.uv.es} \affiliation{\AddrAHEP}

\date{\today}

\begin{abstract}
We propose a realization of the so-called {\it bimodal/schizophrenic} model proposed recently. We assume $S_4$, the permutation group of four objects as flavor symmetry giving tri-bimaximal lepton mixing at leading order. In these models the second massive neutrino state is assumed quasi-Dirac and the remaining neutrinos are Majorana states. In the case of inverse mass hierarchy, the lower bound on the neutrinoless double beta decay  parameter $m_{ee}$ is about two times that of the usual lower bound, within the range of sensitivity of the next generation of experiments.

\end{abstract}

\pacs{
11.30.Hv       
14.60.-z       
14.60.Pq       
14.80.Cp       
14.60.St       
23.40.Bw     
}

\maketitle

\section{Introduction}
Charged particles are Dirac fermions while electrically neutral fermions, like neutrinos,
can be either Dirac or Majorana.
Neutrinoless double beta decay $0\nu\beta\beta$ experiments will confirm (if observed) the Majorana nature of neutrinos~\cite{Schechter:1981bd}. Experiments for $0\nu\beta\beta$ currently under construction will have sensitivity in the range of the inverse hierarchy mass spectrum~\cite{Osipowicz:2001sq,:2008qu,Smolnikov:2008fu,GomezCadenas:2010gs}. Recently, it has been observed in \cite{Allahverdi:2010us} that if the second massive neutrino is of Dirac type
(and so does not participate to the  $0\nu\beta\beta$ decay) in the case of inverse mass hierarchy, the lower bound
on the $0\nu\beta\beta$  parameter $m_{ee}$ is about two times that of the usual bound. In reference \cite{Allahverdi:2010us}, they forbid the Majorana mass for the second neutrino at tree level by means of a flavor symmetry.

The  parameter $m_{ee}$ can be written as combination of neutrino masses, 
namely $m_{ee}=\sum_{i=1}^3 U_{ei}^2m_{\nu_i}$ where $U$ is the lepton mixing matrix.
In the inverse hierarchy case, when three neutrinos are of Majorana type, we have 
\begin{equation}
|m_{ee}|\approx |(\cos^2\theta_{12}+e^{i\alpha}\sin^2\theta_{12})m_{\text{atm}}|>\frac{m_{\text{atm}}}{3}\approx 17\,
\text{meV}.
\end{equation}
If the second massive neutrino is of Dirac type, that is $m_{\nu_2}=0$ in $m_{ee}$ we have
\begin{equation}
|m_{ee}|\approx |\cos^2\theta_{12} m_{\text{atm}}|>\frac{2 m_{\text{atm}}}{3}\approx 34\,
\text{meV}.
\end{equation}
Such a value is in the range of sensitivity of the next generation of experiments and could be ruled out very soon.

A four component spinor $\psi$ is a Majorana spinor if $\psi=\psi^c$ where $\psi^c$ is the charge conjugate
of $\psi$.
The Dirac mass term for a massive spin $1/2$ fermion is given by
\begin{equation}\label{dt}
-m \bar{\psi}\psi
\end{equation}
where $\psi=(\chi,\sigma_2 \phi^*)$ and $\chi, \,\phi$ are two component spinors.
Assuming
$
\chi=\frac{1}{\sqrt{2}}(\rho_2+i \rho_1),\,
\phi=\frac{1}{\sqrt{2}}(\rho_2-i \rho_1),
$
a four component Dirac mass term (\ref{dt}) is equivalent to two Majorana mass terms of equal mass and opposite parity\cite{Nunokawa:2007qh,Langacker:2005pfa}
\begin{equation}
-m \bar{\psi}\psi=-\frac{m}{2}(\rho_1^T\sigma_2\rho_1+\rho_2^T\sigma_2\rho_2).
\end{equation}
For an arbitrary number of Majorana neutrinos the neutrino mass matrix is given by
\begin{equation}
\mathcal{L}=-\frac{1}{2}\sum_{i,j}^n M_{ij}\rho_i^T\sigma_2\rho_j,
\end{equation}
In general the eigenvalues of the mass matrix $M$ can have different signs and we can assign a 
signature matrix $\text{diag}(+,+,...,-,-,..)$. For two neutrino states we can have $\text{diag}(+,-)$
or $\text{diag}(+,+)$. In the former case, if the absolute value of the masses is the same,
the two neutrino types make up a Dirac neutrino. When the two neutrinos
are active-sterile we have the so-called quasi-Dirac neutrino \cite{Valle:1982yw} and when they are active-active
we have the so called pseudo-Dirac neutrino \cite{Wolfenstein:1981kw}.

In Ref.\,\cite{Allahverdi:2010us} the second massive neutrino state has a quasi-Dirac mass\,\footnote{
At leading order $m_{\nu_2}$ is a Dirac state, but at next to leading order it takes a small Majorana mass resulting in a quasi-Dirac state.}, while the first and third
neutrinos get a Majorana mass a la seesaw. Since each flavor state is an admixture of quasi-Dirac and Majorana
states, they call such a case {\it schizophrenic}. For recent studies on this subject see also~\cite{Machado:2010ui,Barry:2010en,Chen:2011dv}. 
There are several models in the literature for exact tri-bimaximal~\cite{Harrison:2002er} based on the group of permutation of four objects $S_4$ as flavor symmetry~\cite{Lam:2008sh,Bazzocchi:2008ej,Ishimori:2008fi,Bazzocchi:2009pv,Grimus:2009pg,Bazzocchi:2009da,Ding:2009iy,Meloni:2009cz,Morisi:2010rk,Adulpravitchai:2010na,Hagedorn:2010th,Ishimori:2010xk,Ishimori:2010fs,Dong:2010zu,Park:2011zt}. Here we study the schizophrenic case assuming the $S_4$ group with extra abelian symmetries as flavor symmetry. Breaking $S_4$ into different $Z_2$ subgroups  respectively in the charged lepton and neutrino sectors we obtain tri-bimaximal mixing at tree-level. The difference between our model and the model of Ref.\,\cite{Allahverdi:2010us} is that they assume the permutation of three objects $S_3$ flavor symmetry instead of $S_4$ and they obtain tri-bimaximal mixing only assuming the charged lepton mass matrix to be diagonal, while in our model the charged lepton mass matrix is diagonal at tree-level by means of $S_4$.\\

The Letter is organized as follow: in section \ref{themodel} we present the model, in section \ref{leptons} we give the 
neutrino and charged lepton mass matrices, in section \ref{vacuum} we study the problem of the vacuum alignments
and we give our conclusions.

\section{The model}\label{themodel}
We extend the Standard Model (SM) with a $G_f=S_4\times Z_3\times Z'_3\times Z''_3$ flavor symmetry where $S_4$
is the permutation group of four objects, $Z_3, Z'_3, Z''_3$ are abelian groups characterized respectively
by ${\omega}^3=1$, ${\omega'}^3=1$ and ${\omega''}^3=1$. 
In order to simplify the study of the $S_4$-alignments of the scalar fields we assume supersymmetry, therefore all the fields are assumed to be superfields.  We also add three right-handed neutrinos and eight scalar isosinglets called flavons.
We assume $\nu_2^c$ to be a singlet of $S_4$ and $\nu_1^c,\nu_3^c$ to form a doublet $\nu_D^c$ of $S_4$. 
The $SU_L(2)$ doublet $L$ and singlet $l^c$ are both triplets $3_1$ of $S_4$.
The matter content of the model is given in table\,\ref{tab1}. 
\begin{table}[h!]
\begin{center}
\begin{tabular}{|c|cccc|c|cccccccc|}
\hline
&$L$&$l^c$ &$\nu_2^c$&$\nu^c_D$&$h^{u,d}$&$\phi_\nu$&$\xi_\nu$&$\varphi_l$&$\chi_l$&$\tilde{\chi}_l$&$\varphi_\nu$&
$\sigma$&$\tilde{\sigma}$\\
\hline
$S_4$&$3_1$&$3_1$&$1_1$&$2$&$1_1$&$3_1$&$1_1$&$2$&$1_1$&$1_1$&2&$1_1$&$1_1$\\
\hline
$Z_3$&1&$\omega^2$&1 &1 &1 &1 &1&$\omega$ &$\omega$&$\omega^2$&1&1&1\\
$Z_3'$&${\omega'}^2$&${\omega'}$&${\omega'}$ &1 &1 &$\omega'$ &${\omega'}^2$&1 &1&1 &1&1&1\\
$Z_3''$&1&1&$1$ &${\omega}''$ &1 &1 &1&1 &1&1 &${\omega''}$&${\omega''}^2$&${\omega''}$
\\
\hline
\end{tabular}
\caption{Matter content of the model.}\label{tab1}
\end{center}
\end{table}\\
The relevant Yukawa terms of the superpotential invariant under $G_f$ are
\begin{equation}
\begin{array}{lll}
  w_l&=&\frac{y_{1l}}{M_\Lambda}Ll^ch^d\chi_l+\frac{y_{2l}}{M_\Lambda}Ll^ch^d\varphi_l,\\
  &&\\
w_\nu&=&\frac{y_{2\nu}}{M_\Lambda^2}L\nu^c_2 h^u \phi_\nu\xi_\nu+\frac{y_{1\nu}}{M_\Lambda^2}L\nu^c_D h^u \phi_\nu\sigma+
y_\sigma\nu_D^c\nu_D^c\tilde{\sigma}+y_{\varphi}\nu_D^c\nu_D^c \varphi_\nu.
\end{array}
\end{equation}
Since $\nu_2^c$ is charged under $Z_3'$ the mass term $\nu_2^c\nu_2^c$ is forbidden.
The scalar flavons take vacuum expectation value (vev) along the following direction of $S_4$ (see section \ref{vacuum})
\begin{equation}\label{allig}
\vev{\phi_\nu}\sim (1,1,1),\quad
\vev{\varphi_\nu}\sim (0,1),\quad
\vev{\varphi_l}\sim (-\sqrt{3},1).
\end{equation}
When the scalar flavons take such vevs, the elements $ST^2,\,S^2TS,\,TS,\,S^3T^2$ leave invariant the charged leptons while the elements 
 $TST,\, TSTS^2,\, S,\, S^3$ leave invariant the neutrino sector. Here $S$ and $T$ are generators of $S_4$, see the Appendix A. 
The different breaking in the charged lepton and neutrino sectors gives (at tree-level) tri-bimaximal mixing.
The scalar $S_4$ singlets $\xi_\nu$, $\chi_l$, $\tilde{\chi}_l$, $\sigma$ and $\tilde{\sigma}$ take vevs different from zero.

\section{mass matrices}\label{leptons}

From the superpotential $w_\nu$ and the vevs alignments given in eq.\,(\ref{allig}) 
the Dirac couplings for the neutrinos are proportional to the following $S_4$ contractions
\begin{eqnarray}
(L\phi)_{1_1}\nu_2^c&\sim& (L_e+L_\mu+L_\tau)\,\nu^c_2\\
(L\phi)_{2}\nu_D^c&\sim& 
\left(
\begin{array}{c} 
\frac{1}{\sqrt{2}}(L_\mu-L_\tau)\\
\frac{1}{\sqrt{6}}(-2L_e+L_\mu+L_\tau)
\end{array}
\right)
\times
\left(
\begin{array}{c} 
\nu_1^c\\
\nu_3^c
\end{array}
\right).
\end{eqnarray}
Then the  Dirac neutrino mass matrix is given by 
\begin{equation}\label{mdirac}
m_D=
\left(
\begin{array}{ccc} 
-\frac{2}{\sqrt{6}}&\frac{1}{\sqrt{3}}&0\\
\frac{1}{\sqrt{6}}&\frac{1}{\sqrt{3}}&\frac{1}{\sqrt{2}}\\
\frac{1}{\sqrt{6}}&\frac{1}{\sqrt{3}}&-\frac{1}{\sqrt{2}}
\end{array}
\right)
\left(
\begin{array}{ccc} 
m_{\nu_1}^D&0&0\\
0&m_{\nu_2}^D&0\\
0&0&m_{\nu_3}^D
\end{array}
\right),
\end{equation}
where 
\begin{equation}\label{mD}
m_{\nu_2}^D=\frac{y_{2\nu}}{M_\Lambda^2 }\vev{h^u}\vev{\phi_\nu}\vev{\xi_\nu},\qquad
m_{\nu_1}^D=m_{\nu_3}^D=\frac{y_{1\nu}}{M_\Lambda^2 }\vev{h^u}\vev{\phi_\nu}\vev{\sigma}.\qquad
\end{equation}
The right-handed Majorana neutrino mass matrix is given by
\begin{equation}
M_R=\left(
\begin{array}{ccc} 
y_\sigma \vev{\tilde{\sigma}} +y_\varphi \vev{\varphi_\nu}& 0&0\\
0&0&0\\
0&0& y_\sigma \vev{\tilde{\sigma}}  -y_\varphi \vev{\varphi_\nu}\\
\end{array}
\right)
\equiv
\left(
\begin{array}{ccc} 
M_1&0&0\\
0&0&0\\
0&0&M_3
\end{array}
\right).
\end{equation}
where $M_1\ne M_3$.
The neutrino mass matrix is diagonalized by the tri-bimaximal mixing matrix, see eq.\,(\ref{mdirac}). 
One neutrino has a quasi-Dirac mass\footnote{Next to leading order terms as well as loop corrections generate a negligible mass term for $\nu_2^c$
then we have a quasi-Dirac state instead of a Dirac one.} \,$m_{\nu_2}\equiv m_{\nu_2}^D$ (see eq.\,(\ref{mD})) and  two neutrinos have Majorana masses 
\begin{equation}
m_{\nu_1}=-\frac{ {m_{\nu_1}^D}^2}{M_1},\quad m_{\nu_3}=-\frac{{m_{\nu_1}^D}^2}{M_3}.
\end{equation}
Note that the masses $m_{\nu_3}$ and $m_{\nu_1}$ are proportional one to each other, so the atmospheric mass spliting arises from the $M_1$ and $M_3$ mass splitting.

Assuming Yukawa couplings of order one and the following value for the scales where the scalar fields take vev  
\begin{equation}\label{scale}
\begin{array}{ccccccccc}
&\vev{h^{u,d}}& < &\vev{\xi_\nu} \sim \vev{\tilde{\sigma}}\sim\vev{\varphi_\nu} &<&\vev{\sigma}\sim\vev{\phi_\nu} \sim \vev{\chi_l}\sim\vev{\varphi_l}\sim\vev{\tilde{\chi}_l}&<&M_\Lambda & \\ \\
\text{scales (GeV)}:\quad& 10^2&,&10^5 &,&10^{13}&,& 10^{15}& 
\end{array}
\end{equation}
then the neutrino masses $m_{\nu_1}$, $m_{\nu_2}$ and $m_{\nu_3}$ are at the eV scale with $M_R\sim 10^5\,GeV$.
As a particular example, taking 
\begin{equation}
y_{1\nu}=0.2200,\quad y_{2\nu}=0.6345,\quad y_{\varphi}=1,\quad y_{\sigma}=-0.2300,\quad
\end{equation}
we have
\begin{equation}
|m_{\nu_1}|=0.0628\,eV ,\quad  |m_{\nu_2}^D|=0.0634\,eV ,\quad  |m_{\nu_3}|= 0.0393\,eV,
\end{equation}
giving about $\Delta m^2_{\text{sol}}\approx 7.5 \cdot 10^{-5} {eV}^2$ and  $\Delta m^2_{\text{atm}}\approx 2.4 \cdot 10^{-3} {eV}^2$
in agreement with data.
We observe that the next to leading order term $\nu^c_2\nu^c_2\xi_\nu^2/M_\Lambda$ is allowed giving
a contribution to $M_R$ of order $10^{-5}\,$GeV that is negligible. 
\\

The charged lepton mass matrix is given from the superpotential $w_l$. It is not difficult to show that
the resulting mass matrix is diagonal. This arises from the $S_4$ symmetry  and the masses are given as\footnote{It is very easy to see that corrections of second order arise by couplings with the flavon $\tilde{\chi}_l$ but those can be reabsorbed in the $y_{1l}$ coupling.}
\begin{eqnarray}
m_{e}&=&\frac{y_{1l} }{M_\Lambda} \vev{h^d}\vev{\chi_l}-\frac{2y_{2l} }{\sqrt{6}M_\Lambda}\vev{h^d} \vev{{\varphi_l}_2}, \\
m_{\mu}&=&\frac{y_{1l} }{M_\Lambda} \vev{h^d}\vev{\chi_l}+\frac{y_{2l} }{M_\Lambda} \vev{h^d}
(\frac{1 }{\sqrt{6}} \vev{{\varphi_l}_2}+ \frac{1 }{\sqrt{2}} \vev{{\varphi_l}_1}),\\
m_{\tau}&=&\frac{y_{1l} }{M_\Lambda} \vev{h^d}\vev{\chi_l} + \frac{y_{2l} }{M_\Lambda} \vev{h^d}
(\frac{1 }{\sqrt{6}} \vev{{\varphi_l}_2}- \frac{1 }{\sqrt{2}} \vev{{\varphi_l}_1}).
\end{eqnarray}
If $ \vev{{\varphi_l}_1}$ and $ \vev{{\varphi_l}_2}$ are free, we have three combinations of 
free parameters and we can fit the charged lepton masses as given below
\begin{eqnarray}
\frac{y_{1l} }{M_\Lambda} \vev{h^d}\vev{\chi_l} &=&  \frac{m_e+m_\mu+m_\tau}{3},\\
\frac{y_{2l} }{M_\Lambda} \vev{h^d} \vev{{\varphi_l}_1}&=&  \frac{m_\mu-m_\tau}{\sqrt{2}}\label{v1},\\
\frac{y_{2l} }{M_\Lambda} \vev{h^d} \vev{{\varphi_l}_2}&=&  \frac{-2m_e+m_\mu+m_\tau}{\sqrt{6}}\label{v2},
\end{eqnarray}
that are of order of the mass of the $\tau$, in agreement with the assumption in eq.\,(\ref{scale}).
In the limit $m_{e,\mu}\to 0$ from eqs.\,(\ref{v1})\,and\,(\ref{v2}) we have
\begin{equation}
\frac{\vev{{\varphi_l}_1}}{\vev{{\varphi_l}_2}}=-\sqrt{3},
\end{equation}
in agreement with the vev alignment given in eq.\,(\ref{allig}). The mass of the muon $m_\mu$ arises from a 
small deviation the alignment $\vev{\varphi_l}\sim(-\sqrt{3}(1+\epsilon),1)$. 
%
Such a deviation can arise from next to leading
order terms in the scalar superpotential as well as by assuming $S_4$ soft breaking terms in the superpotential.
While  the electron mass $m_e$ arises  by means of a fine-tuning of the coupling $y_{1l}$. 
We can easily accommodate the three charged lepton masses in our model, in particular $m_\mu\ll m_\tau$ arises from the alignment $\vev{\varphi_l}\sim(-\sqrt{3},1)$.

\section{vacuum alignments}\label{vacuum}

In the previous sections we showed that assuming the alignments in eq.\,(\ref{allig}) we obtain tri-bimaximal
neutrino mixing and diagonal charged lepton mass matrix. Here we show that the alignment of the flavon fields can arise from the  minimization of the superpotential.
 
The superpotential invariant under $S_4\times Z_3\times Z'_3\times Z''_3$ for the flavon fields of table\,(\ref{tab1})
is given by
\begin{eqnarray}
w&=&\lambda_1\varphi_l\varphi_l\varphi_l+\lambda_2\varphi_l\varphi_l\chi_l+\lambda_3\chi_l\chi_l\chi_l
+\lambda_4\chi_l\tilde{\chi}_l+\lambda_5 \tilde{\chi}_l\tilde{\chi}_l\tilde{\chi}_l+
\nonumber\\
&&\nonumber\\
&+&\,\lambda_6\varphi_\nu\varphi_\nu\varphi_\nu+
\lambda_7\varphi_\nu\varphi_\nu\tilde{\sigma}+\lambda_8\sigma\tilde{\sigma} +
\lambda_9\tilde{\sigma} \tilde{\sigma} \tilde{\sigma} +\lambda_{10}\sigma \sigma \sigma+ 
\nonumber\\
&&\nonumber\\
&+&\,\,\lambda_{11}\phi_\nu\phi_\nu\phi_\nu+\lambda_{12}\xi_\nu\xi_\nu\xi_\nu+\mu_\phi\phi_\nu\phi_\nu+\mu_\xi\xi_\nu\xi_\nu\,,
\label{sistem}
\end{eqnarray}
where the terms proportional to $\mu_\phi$ and $\mu_\xi$ break softly the auxiliary $Z_3'$ symmetry
 while the $Z_3$ and $Z_3''$ are preserved in the superpotential. 
We denote the vevs of the flavon fields as below
\begin{equation}
\begin{array}{c}
\langle \varphi_l \rangle = (u_1,u_2),\quad
\langle \varphi_\nu \rangle = (v_1,v_2),\quad
\langle \phi_\nu \rangle = (r_1,r_2,r_3),\\
\langle \chi_l \rangle = v_\chi ,\quad
\langle \tilde{\chi}_l \rangle = \tilde{v}_\chi ,\quad
\langle \sigma \rangle = v_\sigma ,\quad
\langle \tilde{\sigma} \rangle = \tilde{v}_\sigma ,\quad
\langle \xi_\nu \rangle = v_\xi.
\end{array}
\end{equation}
We show below that $r_1=r_2=r_3=r$, $v_1=0$, $v_2=v$, $u_1=-\sqrt{3} u$ and $u_2=u$ is a possible 
solution of the minimization of the superpotential. 
Then we have to solve the set of equations 
\begin{eqnarray}
\frac{\partial w}{\partial u_1} &=& -\lambda_1 6\sqrt{3} u^2-\lambda_2 2\sqrt{3} u v_\chi=0,
\\
\frac{\partial w}{\partial u_2} &=& \lambda_1 6 u^2+\lambda_2 2 u v_\chi=0,
\\
\frac{\partial w}{\partial v_\chi} &=& \lambda_1 4 u^2+\lambda_3 3 v_\chi^2+\lambda_4 \tilde{v}_\chi=0,
\\
\frac{\partial w}{\partial \tilde{v}_\chi} &=& \lambda_4 v_\chi+\lambda_5 3 \tilde{v}^2_\chi=0,
\\
\frac{\partial w}{\partial v_1} &=& 0,
\\
\frac{\partial w}{\partial v_2} &=& -\lambda_6 3 v^2+\lambda_7 2 v \tilde{v}_\sigma=0,
\\
\frac{\partial w}{\partial v_\sigma} &=& \lambda_{10} 3 v^2_\sigma+\lambda_8 \tilde{v}_\sigma=0,
\\
\frac{\partial w}{\partial \tilde{v}_\sigma} &=&\lambda_7 v^2+\lambda_8  \tilde{v}_\sigma+\lambda_9 3 \tilde{v}^2_\sigma=0, 
\\
\frac{\partial w}{\partial r_1} &=& \lambda_{11} r^2+\mu_\phi 2 r=0,
\\
\frac{\partial w}{\partial r_2} &=& \lambda_{11} r^2+\mu_\phi 2 r=0,
\\
\frac{\partial w}{\partial r_3} &=& \lambda_{11} r^2+\mu_\phi 2 r=0,
\\
\frac{\partial w}{\partial v_\xi} &=& \lambda_{12} 3 {v_\xi}^2+\mu_\xi 2v_\xi =0,
\end{eqnarray}
where we have assumed $r_1=r_2=r_3=r$, $v_1=0$, $v_2=v$, $u_1=-\sqrt{3} u$ and $u_2=u$. 
It is easy to show that such a system admits a solution with $r$, $v$ and  $u$ different from zero
and fixed by the coupling constants of the superpotential in eq.\,(\ref{sistem}).

\bigskip
In summary, we present a realization of the so-called bimodal/schizophrenic ansatz, that is one of the massive neutrino
state is of Dirac-type and the remaining two are Majorana. Then each flavor state is an admixture of Dirac and
Majorana states giving distinct predictions for the neutrinoless double beta decay rate. The model consist of a supersymmetric 
extension of the SM based on the $S_4\times Z_3^3$ flavor symmetry, where we add three right-handed neutrinos, the second of them transforming as a singlet of $S_4$ and 
the other two as a doublet of $S_4$, and eight scalar singlets of the SM. The model also gives tri-bimaximal mixing for neutrinos at leading order. As was pointed out in ~\cite{Allahverdi:2010us} 
this kind of models can be ruled out very soon by neutrinoless double beta decay experiments.

\section{Acknowledgments}

We thank Martin Hirsch for reading the manuscript and helpful comments. This work was supported by the Spanish MICINN under grants FPA2008-00319/FPA and MULTIDARK CSD2009-00064 (Consolider-Ingenio 2010 Programme), by Prometeo/2009/091 (Generalitat Valenciana), by the EU Network grant UNILHC PITN-GA-2009-237920. S. M. is supported by a Juan de la Cierva contract. E. P. is supported by CONACyT (Mexico).

\begin{appendix}

\section{The group $S_4$}
The discrete group $S_4$ is given  by the permutations of four objects and it is composed by 24 elements. 
It can be defined by  two generators  $S$ and $T$ that satisfy 
\begin{equation}\label{rel}
 S^4= T^3= 1,\quad ST^2S=T \,.
 \end{equation}
The 24 elements of $S_4$ belong to five classes
\begin{eqnarray}
\label{classes}
\mathcal{C}_1&:& I \,;\nn\\
\mathcal{C}_2&:& S^2,  T S^2 T^2, S^2 T S^2 T^2 \,; \nn\\
\mathcal{C}_3&:& T,T^2, S^2 T, S^2 T^2,  S T S T^2, S T S, S^2 T S^2,S^3 T S\,;\nn\\
\mathcal{C}_4&:& ST^2,  T^2 S, T S T, T S T S^2, S T S^2, S^2 T S\,;   \nn\\
\mathcal{C}_5&:& S, T S T^2, S T, T S, S^3, S^3 T^2\,.
\end{eqnarray}
The elements of $\mathcal{C}_{2,4} $ define   two different  sets of  $Z_2$ subgroups of $S_4$,   the ones of the  class $\mathcal{C}_{4}$  a set of  $Z_3$ abelian discrete symmetries and those belonging to $\mathcal{C}_{5}$ a set of  $Z_4$ abelian discrete symmetries.
The $S_4$  irreducible representations are two singlets,  $1_1,1_2$, one doublet, $2$,  and two triplets, $3_1$ and $3_2$. We adopt the following basis 
\begin{equation}\label{base2}
S= \left(\begin{array}{cc}- 1&0\\ 0&1  \end{array} \right), \qquad
T=-\frac{1}{2}\left(\begin{array}{cc} 1&\sqrt{3}\\ -\sqrt{3}& 1\end{array} \right)~,
\end{equation}
for the doublet representation and 
\begin{eqnarray}\label{base3}
S_{\pm}= \pm \left(\begin{array}{ccc} -1&0&0\\ 0&0&-1  \\ 0&1  &0\end{array} \right), &\quad& 
T=\left(\begin{array}{ccc} 0&0&1\\ 1&0&0  \\ 0&1  &0\end{array} \right)\,,
\end{eqnarray}
for the triplet representations $3_1$ and $3_2$ respectively. Clearly the generators $(S_+,T)$ and  $(S_-,T)$ define the two triplet representations $3_1,3_2$ respectively.  
All the product rules can be  straightforwardly derived.  We remind the reader to the  product rules reported in \cite{Hagedorn:2006ug} (see also~\cite{Ishimori:2010au}).

The product of $S_4$ representation:

\begin{eqnarray}\nonumber
1_i \times 1_j &=& 1_{\rm
  (i+j) mod 2 +1}\;\;\; \forall \; \rm i \; \mbox{and}
\; \rm j, \\ \nonumber
2 \times 1_i &=& 2  \;\;\; \forall \; \rm i, \\ \nonumber
3_i \times 1_j &=& 3_{\rm
  (i+j) mod 2 +1 }  \;\;\; \forall \; \rm i \; \mbox{and}
\; \rm j,
\end{eqnarray}
\parbox{3in}{
\begin{eqnarray}\nonumber
3_i \times 2 &=&  3_1 +
  3_2  \;\;\; \forall \; \rm i, \\ \nonumber
3_1 \times 3_2 &=& 1_2 + 2 +
3_1 + 3_2,
\end{eqnarray}}
\[
\left[ 2 \times 2\right] = 1_1 + 2 \; ,
\;\;\;\; \left\{ 2 \times 2\right\} = 1_2
\;\;\; \mbox{and} \;\;\;
\left[ 3_i \times 3_i\right] = 1_1 + 2 + 3_1 \; , \;\;\;\; \left\{ 3_i \times 3_i\right\} = 3_2  \;\;\; \forall \; \rm i,
\]
where we introduced the notation $\left[\mu \times \mu \right]$
for the symmetric and $\left\{ \mu \times \mu
\right\}$ for the anti-symmetric part of the product $\mu \times
\mu$. 

\noindent Note that $\nu \times \mu = \mu \times \nu$ for all representations
$\mu$ and $\nu$.
For the irreducible representations:
\[
A \sim 1_1 \;\;\; , \;\;\; B \sim 1_2 \;\;\;
, \;\;\; \left( \begin{array}{cc} a_{1} \\ a_{2} \end{array} \right),
 \left( \begin{array}{cc} a^{\prime}_{1} \\ a^{\prime}_{2} \end{array} \right)
  \sim 2 \; , \;\;
\left( \begin{array}{ccc} b_{1}\\ b_{2} \\ b_{3} \end{array}
\right), \left( \begin{array}{ccc} b_{1} ^{\prime}\\ b^{\prime}_{2} \\ b^{\prime}_{3} \end{array}
\right) \sim 3_1 \;\;\; \mbox{and} \] \[
  \left( \begin{array}{ccc} c_{1}\\ c_{2} \\ c_{3} \end{array}
\right), \left( \begin{array}{ccc} c^{\prime}_{1}\\ c^{\prime}_{2} \\ c^{\prime}_{3} \end{array}
\right)  \sim 3_2 \; .
\]
The explicit products for $1_1$ representation with any $\mu$ representation:
\[
\left( \begin{array}{c} A \, a_1 \\ A \, a_2 \end{array} \right) \sim
2 \;\;\; , \;\;\; \left( \begin{array}{c} A \, b_1 \\ A
    \, b_2 \\ A \, b_3 \end{array} \right) \sim 3_1  \;\;\;
, \;\;\; \left( \begin{array}{c} A \, c_1 \\ A \, c_2 \\ A \, c_3 \end{array} \right) \sim 3_2 \; .
\]
and the product of $1_2$ with the any $\mu$ representation:
\[
\left( \begin{array}{c} -B \, a_2 \\ B \, a_1 \end{array} \right) \sim
2 \;\;\; , \;\;\; \left( \begin{array}{c} B \, b_1 \\ B
    \, b_2 \\ B \, b_3 \end{array} \right) \sim 3_2  \;\;\;
, \;\;\; \left( \begin{array}{c} B \, c_1 \\ B \, c_2 \\ B \, c_3 \end{array} \right) \sim 3_1 \; .
\]
The products of $\mu \times \mu$:\\
\begin{center}\parbox{2.5in}{\begin{center}
 for  $2$
\begin{eqnarray}\nonumber
&a_1 a^{\prime}_1 + a_2 a^{\prime}_2 \sim 1_1,&\\ \nonumber
&-a_1 a^{\prime}_2 + a_2 a^{\prime}_1 \sim 1_2,&\\ \nonumber
&\left( \begin{array}{c} a_1 a^{\prime}_2 + a_2 a^{\prime}_1 \\ a_1 a^{\prime}_1 - a_2 a^{\prime}_2 \end{array} \right) \sim 2,&
\end{eqnarray}
\end{center}
}\end{center}
\parbox{3in}{\begin{center}
for $3_1$
\begin{eqnarray}\nonumber
&\sum \limits _{j=1} ^{3} b_j b^{\prime}_j  \sim 1_1,&\\ \nonumber
&\left( \begin{array}{c} \frac{1}{\sqrt{2}} (b_2 b^{\prime}
_2 - b_3 b^{\prime}_3) \\ \frac{1}{\sqrt{6}} (-2 b_1 b^{\prime}_1 + b_2 b^{\prime}_2 + b_3 b^{\prime}_3) \end{array} \right) \sim 2,& \\ \nonumber
&\left( \begin{array}{c} b_2 b^{\prime}_3 + b_3 b^{\prime}_2 \\ b_1 b^{\prime}_3 + b_3 b^{\prime}_1\\ b_1
    b^{\prime}_2 + b_2 b^{\prime}_1 \end{array} \right) \sim
3_1 \; , \;\; \left(
  \begin{array}{c} b_3 b^{\prime}_2 - b_2 b^{\prime}_3 \\ b_1 b^{\prime}_3 - b_3 b^{\prime}_1 \\ b_2 b^{\prime}_1 -
  b_1 b^{\prime}_2 \end{array} \right) \sim 3_2,&
\end{eqnarray}
\end{center}}
\parbox{3in}{\begin{center}
for $3_2$
\begin{eqnarray} \nonumber
&\sum \limits _{j=1} ^{3} c_j c^{\prime}_j \sim 1_1,&\\ \nonumber
&\left( \begin{array}{c} \frac{1}{\sqrt{2}} (c_2 c^{\prime}
_2 - c_3 c^{\prime}_3) \\ \frac{1}{\sqrt{6}} (-2 c_1 c^{\prime}_1 + c_2 c^{\prime}_2 + c_3 c^{\prime}_3) \end{array} \right) \sim 2,& \\ \nonumber
&\left( \begin{array}{c} c_2 c^{\prime}_3 + c_3 c^{\prime}_2 \\ c_1 c^{\prime}_3 + c_3 c^{\prime}_1\\ c_1
    c^{\prime}_2 + c_2 c^{\prime}_1 \end{array} \right) \sim 3_1\; , \;\;\left(
  \begin{array}{c} c_3 c^{\prime}_2 - c_2 c^{\prime}_3 \\ c_1 c^{\prime}_3 - c_3 c^{\prime}_1 \\ c_2 c^{\prime}_1 -
  c_1 c^{\prime}_2 \end{array} \right) \sim 3_2& .
\end{eqnarray}
\end{center}}\\
For $2 \times 3_1$:
\parbox{2.5in}{
\begin{eqnarray} \nonumber
\left( \begin{array}{c} a_2 b_1 \\ -\frac{1}{2}(\sqrt{3} a_1 b_2 + a_2
    b_2)\\  \frac{1}{2} (\sqrt{3} a_1 b_3 - a_2 b_3) \end{array}
\right) \sim 3_1\\ \nonumber
\left( \begin{array}{c} a_1 b_1 \\ \frac{1}{2}(\sqrt{3} a_2 b_2 - a_1
    b_2)\\  -\frac{1}{2} (\sqrt{3} a_2 b_3 + a_1 b_3) \end{array}
\right) \sim 3_2
\end{eqnarray}}
\parbox{1in}{and for  $2 \times 3_2$}
\parbox{2in}{
\begin{eqnarray} \nonumber
\left( \begin{array}{c} a_1 c_1 \\ \frac{1}{2}(\sqrt{3} a_2 c_2 - a_1
    c_2)\\  -\frac{1}{2} (\sqrt{3} a_2 c_3 + a_1 c_3) \end{array}
\right) \sim 3_1\\ \nonumber
\left( \begin{array}{c} a_2 c_1 \\ -\frac{1}{2}(\sqrt{3} a_1 c_2 + a_2
    c_2)\\  \frac{1}{2} (\sqrt{3} a_1 c_3 - a_2 c_3) \end{array}
\right) \sim 3_2 .
\end{eqnarray}}
\begin{center}
\hspace{-2.8in} 
For $3_1\times 3_2$
\begin{eqnarray} \nonumber
& \sum \limits _{j=1} ^{3} b_j c_j \sim 1_2&\\ \nonumber
&\left( \begin{array}{c}  \frac{1}{\sqrt{6}} (2 b_1 c_1 - b_2 c_2 - b_3
    c_3) \\ \frac{1}{\sqrt{2}} (b_2 c
_2 - b_3 c_3) \end{array} \right) \sim 2& \\ \nonumber
&\left( \begin{array}{c} b_3 c_2 - b_2 c_3 \\ b_1 c_3 - b_3 c_1 \\ b_2 c_1 -
  b_1 c_2 \end{array} \right) \sim 3_1\; , \;\;\left(
  \begin{array}{c} b_2 c_3 + b_3 c_2 \\ b_1 c_3 + b_3 c_1\\ b_1
    c_2 + b_2 c_1 \end{array} \right) \sim 3_2& .
\end{eqnarray}
\end{center}

\end{appendix}

\end{document}